\documentclass[10pt]{amsart}
\usepackage[latin1]{inputenc}
\usepackage[T1]{fontenc}
\usepackage{a4wide}
\usepackage{mathptmx}
\usepackage[colorlinks=true,pdftex]{hyperref}
\usepackage{amsmath}
\usepackage{amsfonts}
\usepackage{amssymb}
\usepackage{graphicx}
\usepackage{mathrsfs}
\usepackage{calrsfs}
\usepackage[abbrev]{amsrefs}
\usepackage{enumerate}
\usepackage{color}

\usepackage{tensind}
\newcommand{\del}{\partial}

\newcommand{\mcA}{\mathcal{A}}

\newcommand{\mcM}{\mathcal{M}}
\newcommand{\mcN}{\mathcal{N}}

\newcommand{\ed}{\operatorname{d}}

\newcommand{\mM}{\mathbb{M}}
\newcommand{\mR}{\mathbb{R}}
\newcommand{\mS}{\mathbb{S}}

\newcommand{\vt}{\vartheta}
\newcommand{\vp}{\varphi}

\newcommand{\cd}{\nabla}       
\newcommand{\ifv}{\Theta}         
\newcommand{\efv}{U}           
\newcommand{\hlf}{\frac{1}{2}} 

\newcommand{\expo}{\operatorname{e}}

\newcommand{\VEg}{\sqrt{|g|}\,\ed^{m+1}x}
\newcommand{\VEgamma}{\sqrt{|\gamma|}\,\ed^{m}x}

\newcommand{\tderiv}[1]{\dot{#1}}
\newcommand{\tderivq}[1]{\dot{#1}^2}
\newcommand{\ttderiv}[1]{\ddot{#1}}

\newcommand{\xderiv}[1]{\del_x {#1}}
\newcommand{\xderivq}[1]{\left ( \del_x {#1} \right )^2}
\newcommand{\xxderiv}[1]{\del_{xx} {#1}}

\newcommand{\yderiv}[1]{\del_y {#1}}
\newcommand{\yderivq}[1]{\left ( \del_y {#1} \right )^2}

\newcommand{\rderiv}[1]{\del_r {#1}}
\newcommand{\rrderiv}[1]{\del_{rr} {#1}}
\newcommand{\rderivq}[1]{\left ( \del_r {#1} \right )^2}

\newcommand{\Lap}[1]{\del_{xx} {#1} - \del_{yy} {#1}} 

\newtheorem{conjecture}{Conjecture}

\tensordelimiter{?}

\begin{document}
  \title{Free Versus Constrained Evolution of the 2+1 Equivariant Wave Map}
  
  \author{Ralf Peter}
  \author{J\"org Frauendiener}
  
  \address{Department for Mathematics and Statistics\\ University of 
     Otago\\ PO Box 56, Dunedin 9054, New Zealand}
  
  \thanks{We thank Christian Lubich for pointing out the Rattle method to 
     us. RP was supported by a DAAD and a University of Otago Doctoral 
     Scholarship. This research was partially funded by the 
     Royal Society of New Zealand via the Marsden Fund.}
  
  \subjclass[2010]{35L67, 35L70, 65M20, 65P10, 74H35}
  
  \email{rpeter@maths.otago.ac.nz}
  \email{joergf@maths.otago.ac.nz}
  \begin{abstract}
    We compare the numerical solutions of the 2+1 equivariant Wave Map
    problem computed with the symplectic, constraint respecting Rattle
    algorithm and the well known fourth order Runge-Kutta method. We
    show the advantages of the Rattle algorithm for constrained system
    compared to the free evolution with the Runge-Kutta method. We
    also present an expression, which represents the energy loss due to
    constraint violation. Taking this expression into account we can
    achieve energy conservation for the Runge-Kutta scheme, which is
    better than with the Rattle method. Using the symplectic scheme
    with constraint enforcement we can reproduce previous calculations
    of the equivariant case without imposing the symmetry
    explicitly, thereby confirming that the critical behaviour is
    stable.
  \end{abstract}
  \maketitle
  \section{Introduction}
  Down to the present day, the use of symplectic, or more general geometric, 
  numerical integration methods in the field of (general) relativistic (field) 
  equations is not very common. To our knowledge the first work was done in 
  \cite{BM1993} and \cite{BGS1997} to solve ODEs in the context of cosmological 
  space-times. The step to field equations was done in \cite{RL2008} and \cite
  {Richter2009}. A symplectic integrator for post Newtonian equations was 
  developed only recently \cite{LWB2010}.  A reason why this has not been done 
  in the past very often, is the different symplectic structure of General 
  Relativity compared to usual Hamiltonian systems \cite{Frauendiener2008}.

  
  Here, we want to present another contribution to the problem of 
  solving relativistic field equations with the help of symplectic integrators.
  We compare the numerical solution of a 2+1 dimensional Wave Map 
  system solved with a symplectic, constraint preserving (at least in a given 
  accuracy) integration scheme with the standard fourth order Runge-Kutta 
  method. We will show that we can reproduce the qualitative results gained by 
  Bizo\'n et.al. \cite{Bizon2001} and Isenberg and Liebling \cite{IL2002} in 
  the so called equivariant case.
  
  The outline of this article is the following: in section \ref{sec:Rattle} we 
  give a brief description of the symplectic integration method, 
  which we will use for our numerical studies. As already mentioned, we 
  solve a Wave Map system to show the differences between the integration 
  methods. So in section \ref{sec:Wave_Map} we describe Wave Maps in 
  general and concentrate then on the 2+1 equivariant case. In section \ref
  {sec:flux} we derive a correction term for the energy, which is caused by the 
  constraint violation during the numerical time evolution. The numerical 
  setup, including the choice of the initial data and the boundary conditions 
  are presented in section \ref{sec:numerics}. Finally, in section \ref
  {sec:results}, we present some results of our numerical studies.
  
  Our numerical results can be put into three  independent 
  categories:
  \begin{enumerate}[(a)]
    \item The comparison of the standard fourth order Runge-Kutta method with 
      the Rattle method.
    \item The correction of the energy by taking the constraint violation into 
      account.
    \item The blow up dynamics of the 2+1 equivariant Wave Map, where we 
      reproduce known results.
  \end{enumerate}
  \section{The Rattle Method}\label{sec:Rattle}
  The Rattle method is a numerical integration method for constrained
  ODE systems. It is a further development of the Shake algorithm
  \cite{RCB1977} (which itself is based on the well known
  St\"ormer-Verlet method \cite{HLW2003}) to take the presence of constraints 
  into account. It was developed by Andersen~ \cite{Andersen1983} in the
  context of molecular dynamics. Later, it was shown by Leimkuhler and
  Skeel \cite{LeimkuhlerSkeel1994} that the Rattle method belongs to
  the group of the symplectic integrators.

  Symplectic integrators are a certain class of geometric integrators
  \cites {HLW2003,LeimkuhlerReich,HLW2006,McLQ2006}. Their
  characteristic feature is the conservation of a geometric structure
  or quantity. Such quantities can be for example the energy, the
  momentum, the phase space volume or, like in the case of the
  symplectic integrators, the symplectic 2-form $ \omega = \ed q^a
  \wedge \ed p_a$.
  
  In many branches of physics and applied mathematics, including
  celestial mechanics, molecular dynamics and quantum mechanics
  symplectic integrators have been used very successfully for
  years. Very often the numerical results, especially in terms of long-term
  energy and (angular) momentum conservation, are much better compared
  to non-symplectic integrators \cite{HLW2006}.
  
  Now we give a brief description of the Rattle method. For further 
  details we refer to the original paper \cite{Andersen1983}. We consider a 
  Lagrangian system with holonomic constraints $\phi^A$ and an action of the 
  following form\footnote{It is only possible for systems with holonomic 
  constraints to write the action in this form.}:
  \begin{align*}
    \mcA[q,\dot q, \lambda] = \int \left ( \hlf\, M_{kj} \tderiv{q}^k\tderiv{q}
    ^j - V(q) + 
    \lambda_A\phi^A(q) \right ) \ed t
  \end{align*}
  which leads to the Euler-Lagrange equations
  \begin{align}\nonumber
    \frac{d}{dt}\left(M_{jk}\dot{q}^k\right) + \del_j V - \lambda_A \, \del_j 
    \phi^A &= 0,\\\label{eq:Rattle_Constr}
    \phi^A(q) &= 0\,.
  \end{align}
  The constraint equation \eqref{eq:Rattle_Constr} implies also an additional 
  constraint on the velocities:
  \begin{align*}
    \psi^A (q,\tderiv{q}) := \frac{d}{dt}{\phi}^A(q) = \tderiv{q}^j\,\del_j 
    \phi^A = 0\, .
  \end{align*}
  With $M$ we denote the mass matrix, with $q$ the generalized coordinates, $
  \tderiv{q}$ their corresponding velocities, $\lambda_A$ are the Lagrangian 
  multipliers and $V$ is the potential. With the small Latin indices we label 
  the dynamical variables and with the capital indices the constraints. We use 
  here and throughout the whole article the common dot notation for time 
  derivatives: $\dot{f} := \del_t f$.
  
  A time-step with the Rattle method starts with a St\"ormer-Verlet
  step using the unconstrained equations. In general the result does
  not satisfy the constraints. Therefore, the next step is to compute
  the appropriate amount of the constraint force $- \lambda_A \,
  \del_j \phi^A$, which is necessary to push the estimated solution to
  the constraint surface. This is done by first correcting the
  coordinates $q^j$ by iteratively solving the constraint $\phi^A = 0$
  to determine the values of the Lagrange multipliers. Once these
  constraints are satisfied to a given accuracy the velocity
  constraints $\psi^A=0$ are solved in a similar way to correct the
  velocities.
  \section{Wave Maps}\label{sec:Wave_Map}
  Wave Maps are systems of generalized wave equations. They appear in several 
  physical models, for example as nonlinear sigma models in particle physics 
  \cite{Ketov2009} as well as the Einstein equations for a certain class of 
  electro-vacuum space-times \cite{BCM1995} or in cosmological models 
  \cites{Ringstroem2004}. On the other hand, Wave Maps are of 
  mathematical interest, because of their nonlinear structure and the 
  possibility for singularity formation. Wave Maps are also the simplest 
  example for geometric wave equations \cite{ShatahStruwe}.
  \subsection{General}
  In general, a Wave Map describes a mapping from the $m+1$-dimensional
  Lorentzian manifold $\mcM$ into a $n$-dimensional Riemannian manifold $\mcN$. 
  Usually we distinguish between two different kinds of formulations for the 
  Wave Maps. The first, so-called intrinsic formulation, describes the mapping 
  directly from the base manifold  $\mcM$ into the target manifold $\mcN$.
  \begin{align*}
    \ifv: \quad \mcM &\rightarrow \mcN\\
                   x &\mapsto \xi = \ifv(x)\,.
  \end{align*}
  We denote here and in the following the coordinates in $\mcM$ with $x = 
  (x^0,x^1,...,x^m)$ and with $\xi = (\xi^1,...,\xi^n)$ the coordinates in $
  \mcN$. Further, $g_{ab}$ is the metric on $\mcM$ and $h_{\beta\gamma}$ the 
  metric on $\mcN$, so small Latin indices refer to manifold $\mcM$ and small 
  Greek indices to $\mcN$.
  
  The action for the intrinsic formulation of the Wave Map is
  \begin{align*}
    \mcA[\ifv,\nabla\ifv] = \int_{\mcM} g^{ab}\, \cd_a \ifv^{\beta}\, \cd_b 
    \ifv^{\gamma}\, h_{\beta\gamma}\VEg\,.
  \end{align*}
  We denote with $\cd_a$ the covariant derivative with respect to the 
  metric $g_{ab}$ on $\mcM$ and with $g$ the determinant of $g_{ab}$.
  
  The variation of the action with respect to the field variables
  $\ifv^{\beta}$ gives us the equations of motion:
  \begin{align}\label{intrinsic_eom}
    \square_g \ifv^{\beta}
    + ?\Gamma^{\beta}_{\mu\nu}?\,\cd^a\ifv^{\mu}\, \cd_a\ifv^{\nu} = 0\,.
  \end{align}
  With $\square_g = \nabla^a\nabla_a$ we denote the d'Alembert operator 
  (wave operator) on the base manifold $\mcM$ and with $?
  \Gamma^{\beta}_{\mu\nu}? = ?\Gamma^{\beta}_{\mu\nu}? (\ifv(x))$ 
  the Christoffel symbols corresponding to the metric $h_{\beta\gamma}$ 
  on $\mcN$. Equation \eqref{intrinsic_eom} shows clearly that Wave Maps 
  are generalizations of the wave equation ($\mcN = \mR$) on the one side 
  and the geodesic equation ($\mcM = \mR$) on the other side.
  
  A different way to describe Wave Maps is the so-called extrinsic 
  formulation. In this formulation we assume the target manifold $\mcN$ 
  isometrically embedded into a higher dimensional euclidean space 
  $\mR^p$, $p > n$.
  \begin{align*}
    \efv: \quad \mcM &\rightarrow \mcN \hookrightarrow \mR^p\\
                   x &\mapsto z = \efv(x)
  \end{align*}
  with $z = z^1,...,z^p$ the coordinates on $\mR^p$ and the notation for the 
  variables in $\mcM$
  being the same as for the intrinsic formulation.  In this case we
  need additional conditions $\phi(z)=0$, which ensure that the image
  of the mapping lies in the sub-manifold $\mcN$ of $\mR^p$. We can do
  that by attaching the constraint expressions $\phi(z)$ to the action
  via Lagrangian multipliers. For the following purpose we can assume
  that it is enough that $p = n + 1$, which means that we need only
  one equation to describe the embedding of $\mcN$ into
  $\mR^{n+1}$. We write for the action of the extrinsic formulation
  \begin{align}\label{extrinsic_action}
    \mcA[\efv,\nabla \efv,\lambda]= \int_{\mcM}\left ( g^{ab}\, \cd_a \efv^A\, 
    \cd_b\,\efv^B\, \delta_{AB} + 2\lambda\phi \right )\VEg\,.
  \end{align}
  The capital Latin indices refer to $\mR^{n+1}$ and $\delta_
  {AB}$ denotes the Euclidean metric on $\mR^{n+1}$.
  
  Again, extremization of this action results in the equations of 
  motion for this formulation:
  \begin{align*}
    \square_g \efv^A + \lambda\,\del^A \phi &= 0,\\
    \phi(z) &= 0.
  \end{align*}
  \subsection{2+1 Wave Map}
  In the following we will concentrate on a Wave Map with the base 
  manifold $\mcM = \mM^{2+1}$, the $2+1$ dimensional Minkowski space-time and 
  the target manifold $\mcN = \mS^2$, the 2-sphere.
  
  \paragraph{\textbf{Intrinsic Formulation}} We will use the 
  intrinsic formulation to discuss the basic properties of the 2+1 Wave Map. In 
  this formulation, we write for the map:
  \begin{align*}
    \Theta: \quad \mM^{2+1} = \mR^2 \times \mR &\rightarrow \mS^2\\
          (x^0,x^1,x^2) &\mapsto \left ( \vt(x^0,x^1,x^2), 
          \varphi(x^0,x^1,x^2) \right )\,.
  \end{align*}
  With the common notation for spherical coordinates on the 2-sphere we write 
  for the line element
  \begin{align*}
    \ed s_{\mS^2}^2 = \ed\vt^2 + \sin^2\vt\, \ed\varphi^2\, .
  \end{align*}
  The equations of motion take the following form:
  \begin{align}\label{eq:intr_theta}
    \square_g \vt - \sin\vt\cos\vt\, \cd^a \vp \cd_a \vp &= 
    0,\\\label{eq:intr_phi}
    \square_g \vp + 2\cot\vt \, \cd^a \vt \cd_a \vp &= 0\,.
  \end{align}
  Further, we choose polar coordinates on the Minkowski space-time with the 
  metric
  \begin{align*}
    \ed s_{\mM^{2+1}}^2 = g_{ab}\,\ed x^a \ed x^b = \ed t^2 - \ed r^2 - r^2 
    \ed \sigma^2
  \end{align*}
  where $r$ is the radial and $\sigma$ the angular coordinate. The
  most commonly studied situation for this Wave Map is the so-called
  equivariant case. In this case one sets $\vt(t,r,\sigma) = 
  \vt(t,r)$ and $\vp(t,r,\sigma) = \sigma$. The latter 
  assumption means that the Wave Map maps rotations on Minkowski
  space around the origin to rotations of the sphere around the 3-axis
  with the same angle. In this case, the equation \eqref{eq:intr_phi}
  is identically satisfied and \eqref{eq:intr_theta} reduces to
  \begin{align}\label{eq:equiv_wm}
    \ttderiv{\vt} - \rrderiv{\vt} - \frac{1}
    {r}\,\rderiv{\vt} + \frac{\sin(2\vt)}{2r^2} = 0\,.
  \end{align}
  By confining to the equivariant case one needs to solve only one
  equation, but at the cost of having to deal with a coordinate
  singularity at $r = 0$. This equation together with the initial data
  \begin{align}\label{eq:intr_initial_data}
    \vt(0,r) = \vt_0(r) \qquad \text{and} \qquad \tderiv{\vt}(0,r) = \vt_1(r)
  \end{align}
  is the Cauchy problem, which was studied in \cite{Bizon2001}. Equation \eqref
  {eq:equiv_wm} has two non-trivial, static solutions with $\vt(r) \in [0,\pi]
  $:
  \begin{align}\label{eq:intr_static_sol}
    \vt_{\text{S}}(r) = 2\arctan(r^{\pm1}).
  \end{align}

  The local well posedness of the Cauchy problem for Wave Maps 
  with general target manifold, was proven in \cite{KlaMach1993} 
  and \cite{KlaSel1997}. The study of the global well posedness started with 
  the investigation of the equivariant case. This was 
  done in \cite{ChrTah-Zad1993_1,ChrTah-Zad1993_2,ShaTah-
  Zad1994,Struwe2002,Struwe2003_1}. The problem of global well-posedness for 
  the non-symmetric 2+1 Wave Map with small initial data\footnote{\textit
  {Small} means here, small with respect to the energy. In other words, the 
  energy is below a critical value.} was first addressed by Tataru \cite
  {Tataru1998}. Later work include \cite
  {Tao2001a,Tao2001b,KlaRod2001,ShaStr2002,Tataru2005}. For the case of a 
  hyperbolic target manifold, see \cite{Krieger2003,Krieger2004}. Just recently 
  the global well-posedness for the Cauchy problem in 2+1 dimensions for large 
  energy initial data was first proven by Krieger and Schlag for the hyperbolic 
  target \cite{KriSch2009} and by Sterbenz and Tataru for the $\mS^2$ case 
  \cite{SterbenzTataru2010_1,SterbenzTataru2010_2}.
  Finally, we want to recommend here the review article by Krieger \cite
  {Krieger2008} on the Wave Map problem which covers the most important results 
  and references up to the year 2008.

  The main interest in studying the Wave Map equations and
  particularly the 2+1 Wave Map into a 2-sphere is the formation of
  singularities. These question is closely related to the 
  previously discussed problem of global well-posedness or regularity. In the 
  1+1 case it is easy to show that no
  singularities can form, because the Wave Map equations are
  equivalent to the standard 1+1 wave equation. For spatial dimensions larger 
  than two it is known that
  the Wave Maps can form singularities. See \cite {Tataru2004,Krieger2008} for
  detailed information on this. Only for the 2+1 case it was for a
  long time not known. In 2001 Bizo\'n et. al. presented the 
  first crucial numerical evidence for the singularity formation of the 2+1 
  dimensional Wave Map into the 2-sphere. We are going to present their results 
  and point out the important contributions during the last years in 
  this exciting field. Based on their numerical observations of the
  equivariant case Bizo\'n et. al. formulated three conjectures about
  the singularity formation:
  
  \begin{conjecture}[On blow-up for large data]\label{conj:Bizon_1}
    For initial data \eqref{eq:intr_initial_data} with sufficiently large 
    energy, the solutions of equation \eqref{eq:equiv_wm} blow up in finite 
    time 
    in the sense that the derivative $\del_r \vt(t,0)$ diverges as  $t \nearrow 
    T$ for some $T > 0$.
  \end{conjecture}
  
  \begin{conjecture}[On blow-up profile]\label{conj:Bizon_2}
    Suppose that the solution $\vt(t,r)$ of the initial-value problem \eqref
    {eq:equiv_wm} and \eqref{eq:intr_initial_data} blows up at some time $T 
    > 0$. Then, there exists a positive function $s(t) \searrow 0$ for $t 
    \nearrow T$ such that
    \begin{align*}
      \lim_{t \nearrow T} \vt(t,s(t)r) = \vt_{\text{S}}(r)
    \end{align*}
  \end{conjecture}
  
  This conjecture was proven by Struwe in \cite
  {Struwe2003_2}, where, he not only proved the above statement, but further he 
  showed that the existence of a non-trivial (non-constant) harmonic map is a 
  necessary condition for a singularity formation. Struwe showed that the 
  singularity formation appears as an energy concentration at the origin and 
  the \textit{bubbling-off of a harmonic map}. The latter is the decomposition 
  of the solution near the blow 
  up time $T$ in the following form (see \cite{KST2008})
  \begin{align}\label{eq:blow_up_decomposition}
    \vt(t,r) = \vt_{\text{S}}\left(s(t)r\right) + R(t,r)\,.
  \end{align}
  The function $R(t,r)$ is a radiative term and $\vt_{\text{S}}\left(s
  (t)r\right)$ the rescaled static solution. The error term's local energy
  goes to zero as the time reaches the 
  blow up time $T$. Equipped with the result in \cite{Struwe2003_2} Krieger, 
  Schlag and Tataru were able to construct the first blow up scenario for the 
  2+1 Wave Map into the 2-sphere \cite{KST2008}. This was later generalized to 
  surfaces of revolution, which includes the 2-sphere, as target manifold.
  
  \begin{conjecture}[On energy concentration]\label{conj:Bizon_3}
    Suppose that the solution $\vt(t,r)$ of the initial-value problem \eqref
    {eq:equiv_wm} and \eqref{eq:intr_initial_data} blows up at some time $T 
    > 0$. Define the kinetic and the potential energies at time $t < T$ inside 
    the past light-cone of the singularity by
    \begin{align*}
      E_{\text{kin}}(t) = \pi \int_0^{T-t}\tderivq{\vt} r\,\ed r
      \quad \text{and} \quad
      E_{\text{pot}}(t) = \pi \int_0^{T-t}\left[ \rderivq{\vt} + \frac{\sin^2 
      \vt}{r^2}\right] r\,\ed r
    \end{align*}
    Then:
    \begin{enumerate}[(a)]
      \item the kinetic energy tends to zero at the singularity
        \begin{align*}
          \lim_{t \nearrow T} E_{\text{kin}}(t) = 0
        \end{align*}
      \item the potential energy equal to the energy of the static solution $
      \vt_{\text{S}}$ concentrates at the singularity
      \begin{align*}
        \lim_{t \nearrow T} E_{\text{pot}}(t) = E[\vt_{\text{S}}] = 4\pi\,.
      \end{align*}
    \end{enumerate}
  \end{conjecture}
  
  The behaviour of the energy, as described in the last 
  conjecture is a direct consequence of the splitting \eqref
  {eq:blow_up_decomposition} if one considers the vanishing energy of the 
  radiation term in the past light cone.
  
  In a recent detailed study on the singularity formation 
  Rapha\"el and 
  Rodnianski constructed a stable blow up scenario \cite{RaphaelRodnianski}. A 
  different approach by Ovchinnikov and Sigal can be found in \cite
  {OvchinnikovSigal}. In both publications the 
  authors give an analytical form for the scaling function $s(t)$. We are going 
  to use this result later on in section \ref{sec:results_blowup}. We want to 
  mention here that the results by Rapha\"el and Rodnianski cover all homotopy 
  classes of the equivariant Wave Map. The singularity formation for higher 
  homotopy classes was already studied by Rodnianski and Sterbenz in \cite
  {RS2008}.
  
  \paragraph{\textbf{Extrinsic Formulation}} In the later 
  numerical 
  simulations, we will use the extrinsic formulation, where we use Cartesian 
  coordinates in the Euclidean space to avoid coordinate singularities. We 
  write for the Wave Map:
  \begin{align*}
    \efv: \quad \mM^{2+1} &\rightarrow \mS^2 \hookrightarrow \mR^3\\
          x &\mapsto \left ( \efv^1(t,x,y), \efv^2(t,x,y), \efv^3(t,x,y) 
          \right )\, .
  \end{align*}
  The line element simply is
  \begin{align*}
    \ed s_{\mR^3}^2 = (\ed \efv^1)^2 + (\ed \efv^2)^2 + (\ed \efv^3)^2\,.
  \end{align*}
  The field equations, consisting of the equations of motion and the constraint 
  equation, take the form:
  \begin{align}\label{eq:extr_eom}
    \square_g \efv^A - 2\lambda \efv^A &= 0,\\\label{eq:extr_constr}
    \efv^A \efv_A - 1 &= 0\,.
  \end{align}
  The Lagrangian parameter $\lambda = \lambda(t,x,y)$ can be computed by 
  multiplying the equation of motion \eqref{eq:extr_eom} with $u_A$ and the use 
  of the constraint equation \eqref{eq:extr_constr} and its derivative. This 
  leads to
  \begin{align*}
    \lambda = -\frac{1}{2}\left ( \cd^a \efv^B \cd_a \efv_B \right )\,.
  \end{align*}
  The resulting non-linear wave equation has the constraint condition already 
  imprinted
  \begin{align*}
    \square_g \efv^A + \left ( \cd^a \efv^B \cd_a \efv_B \right ) \efv^A &= 
    0\,.
  \end{align*}
  Now, we also introduce Cartesian coordinates on the base manifold $\mM^{2+1}$
  \begin{align*}
    \ed s_{\mM^{2+1}}^2 = \ed t^2 - \ed x^2 - \ed y^2\,.
  \end{align*}
  By the relabelling $u:=\efv^1$, $v:=\efv^2$ and $w:=\efv^3$ we can write the 
  equations \eqref{eq:extr_eom} and \eqref{eq:extr_constr} in the explicit form
  \begin{align}\label{eq:extr_eom_u}
    \ttderiv{u} - \Lap{u} - 2\lambda u &= 0,\\\label{eq:extr_eom_v}
    \ttderiv{v} - \Lap{v} - 2\lambda v &= 0,\\\label{eq:extr_eom_w}
    \ttderiv{w} - \Lap{w} - 2\lambda w &= 0,\\\label{eq:extr_constr_uvw}
    \phi(u,v,w) = u^2 + v^2 + w^2 - 1 &\equiv 0.
  \end{align}
  The Lagrangian multiplier is given explicitly by
  \begin{align}\label{eq:lag_mult_uvw}
  \begin{split}
    \lambda = -\hlf\left[\tderivq{u} + \tderivq{v} +\tderivq{w}
        - \xderivq{u} - \xderivq{v} - \xderivq{w}
        - \yderivq{u} - \yderivq{v} - \yderivq{w}\right]
  \end{split}
  \end{align}
  The hidden constraint for the velocities is the time derivative of \eqref
  {eq:extr_constr_uvw}:
  \begin{align}\label{eq:extr_velo_constr_uvw}
    \psi(u,v,w,\dot u,\dot v,\dot w) = 2u\tderiv{u} + 2v\tderiv{v} + 2w\tderiv
    {w} \equiv 0\,.
  \end{align}
  
  We use the following relation between the intrinsic and extrinsic 
  formulation:
  \begin{align}\label{eq:trafo_intr_extr_u}
    u &= \sin \vt \cos \vp 
         \\\label{eq:trafo_intr_extr_v}
    v &= \sin \vt \sin \vp
         \\\label{eq:trafo_intr_extr_w}
    w &= \cos \vt\,.
  \end{align}
  Now, we can use these 
  relations to write the static solutions \eqref{eq:intr_static_sol} in the 
  extrinsic formulation
  \begin{align}\label{eq:extr_static_sol}
  \begin{split}
    u_{\text{S}}(x,y) = \frac{2x}{1 + r^2},\qquad
    v_{\text{S}}(x,y) = \frac{2y}{1 + r^2},\qquad
    w_{\text{S}}(x,y) = w_{\text{S}}(r) = \pm\frac{1 - r^2}{1 + r^2}\,,
  \end{split}
  \end{align}
  with the radial coordinate $r = \sqrt{x^2 + y^2}$. The $\pm$ sign in the 
  expression for $w_{\text{S}}(x,y) $ corresponds to the 
  $\pm$ sign in \eqref{eq:intr_static_sol}. It is easy to see that the static 
  solutions, which are maps from $\mathbb{R}^2$ into $S^2$, are exactly the 
  stereographic projections from the south ($+$) and the north ($-$) poles, 
  respectively.

  In conjecture \ref{conj:Bizon_3} the existence of a scaling function
  $s(t)$ is mentioned with the property to rescale the static solution
  in a way to approximate the dynamical solution around the origin. We
  can determine the scale factor at every fixed time as follows: we
  choose the component function $w(t,r)$ of the Wave Map and choose
  $s(t)$ in such a way that the second radial derivatives at the
  origin of the static and the rescaled dynamical solution \eqref
  {eq:extr_static_sol} agree\footnote{In \cite{Bizon2001} the
    first derivatives are used to compute the scaling function. But in the 
    extrinsic formulation the function $w_{\text{S}}$ of the static solution is 
    an even function and thus has vanishing odd derivatives at the origin.}. 
  We use the function
  $w_{\text{S}}(r)$ with the $+$ sign because we prescribe initial
  data in such a way that the stereographic projection from the south
  pole is singled out. The scaling relationship
  \begin{align*}
    w_{\text{S}}(r) = w(t, s(t)r)
  \end{align*}
  is assumed to be valid in a neighbourhood of the origin.
  Taking the second radial derivative and evaluating at the origin
  gives
  \begin{align*}
    w_{\text{S}}''(0) = s(t)^2\,\del_{rr} w(t,0),
  \end{align*}
  from which we can determine the scaling factor as
  \begin{equation}
    s(t) = \frac{2}{\sqrt{\left |\del_{rr} w(t,0)\right |}}\,.\label
    {eq:scaling_func}
  \end{equation}
  \section{Energy Conservation}\label{sec:flux}
  From \eqref{extrinsic_action} we can get the energy-momentum tensor by 
  variation of the action with respect to the inverse metric $g^{ab}$
  of the base space 
  $\mcM
  $:
  \begin{align}\label{eq:EMT_extr}
    T_{ab} = \cd_a \efv^A \cd_b \efv^B \delta_{AB} - \hlf \cd_c \efv^A \cd^c 
    \efv^B \delta_{AB}\, g_{ab} - \lambda \phi g_{ab}
  \end{align}
  
  In general, the energy of a matter field on a spatial slice in
  $\mcM$ as measured by an observer field with 4-velocity $t^a$ is defined by
  \begin{align*}
    E(t) = \hlf \int_{\Sigma_t} t^a n^b T_{ab}\,\VEgamma\,.
  \end{align*}
  The slicing of the manifold $\mcM$ implies a (negative definite)
  spatial metric $\gamma_{ab}$ on the slice $\Sigma_t$, which has the
  future-pointing normal vector field $n^b$. For our purposes it is
  enough to consider flat space-like hyper-planes
  $\Sigma_t$. Furthermore, we choose static observers so that $t^a =
  (1,0,0) = n^a$. This and the energy-momentum tensor
  \eqref{eq:EMT_extr} lead to the following form of the energy
  contained inside a spatial domain $\Omega\subseteq \mR^2$
  \begin{align*}
    E_\phi(t) = \hlf \int_{\Omega} \left[ \tderiv{\efv}^A \tderiv{\efv}_A
    + \xderiv{\efv^A} \xderiv{\efv_A} + \yderiv{\efv^A} \yderiv{\efv_A} -
    \lambda \phi(\efv) \right]\ed x\,\ed y .
  \end{align*}
  Here we write $E_\phi$ in order to indicate that we do not
  necessarily impose the constraint. This has the consequence that
  the Energy-Momentum tensor is not divergence-free, as we find by
  computing its divergence
  \begin{align}\label{eq:EMT_rest}
    \cd^a T_{ab} = - \left ( \del_b \lambda \right ) \phi\, .
  \end{align}
  From \eqref{eq:EMT_rest} we can see clearly that the energy-momentum tensor 
  is divergence free, provided $\phi = 0$, i.e., if and only if the
  constraint is satisfied. But what will happen if this is not the case? 
  This is what almost invariably happens in a numerical simulation. We
  can use the above 
  formula to find the energy loss due to the constraint 
  violation.
  
  We need the projection of \eqref{eq:EMT_rest} along the (constant)
  time-like vector $t^a$
  \begin{align}\label{eq:cont_eq}
    t^b\nabla^a T_{ab} = - \left ( t^b\del_b \lambda \right ) \phi\, .
  \end{align}
  We can interpret this equation as a continuity equation with an additional 
  source on the right hand side. Using the fact that we are working on 
  Minkowski space-time in Cartesian coordinates, we write for \eqref
  {eq:cont_eq} with a dot as the time derivative
  \begin{equation}
    \label{eq:deviation_1}
    \tderiv{T}_{00} + \del^k T_{0k} = - \tderiv{\lambda}\phi
  \end{equation}
  where small Latin indices from the middle of the alphabet denote the spatial 
  coordinates of the Minkowski spacetime.
  
  The required components of the energy-momentum tensor are
  \begin{align}\nonumber
    T_{00} &= \hlf\,\tderiv{\efv}^A\tderiv{\efv}_A - \hlf\,\del^k \efv^A \del_k 
    \efv_A - \lambda
    \phi =: \rho - \lambda\phi\\\label{eq:momentum_flux_density}
    T_{0k} &= \tderiv{\efv}^A \del_k \efv_A =: j_k
  \end{align}
  with the `true' energy density $\rho$ of the system when the constraint is
  satisfied and the momentum  
  current density $j_k$. Now we can express \eqref{eq:deviation_1} as
  \begin{align*}
    \tderiv{\rho} - \tderiv{\lambda}\phi - \lambda\tderiv{\phi} + \del_k j^k &=
    - \tderiv{\lambda}\phi\\
   \implies \qquad \tderiv{\rho} + \del_k j^k &= \lambda\tderiv{\phi}
  \end{align*}
  and integrate over the domain $\Omega$ at a fixed time
  \begin{align*}
    \int_{\Omega} \tderiv{\rho}\, \ed x\,\ed y + \int_{\Omega} \del_k j^k\, \ed 
    x\,\ed y = 
    \int_{\Omega} \lambda\tderiv{\phi}\, \ed x\,\ed y\,.
  \end{align*}
  The first term on the left hand side in this equation is the time derivative 
  of the `true' energy $E$ and the second term is an integral over the 
  divergence of $j_k$. With the help of Gauss' theorem, we can write the 
  divergence term as a surface integral over the boundary $\del \Omega$
  \begin{align}\nonumber
    \tderiv{E} + \int_{\del \Omega} n_k j^k\,\ed S &= \int_{\Omega} \lambda
    \tderiv{\phi}\, \ed x\,\ed y\\\label{eq:energy_deviation_2}
    \implies\tderiv{E} &= \int_{\Omega} \lambda \psi\, \ed x\,\ed y
    - \int_{\del \Omega} n_k j^k\,\ed S
  \end{align}
  with the normal vector field $n_k$ on the boundary and the boundary element $
  \ed S$. The change of the energy between two instants of time is obtained by
  integrating  over a time interval $[t_1,t_2]$
  \begin{align}\label{eq:deviation_2}
    \Delta E = E(t_2) - E(t_1) = \int_{t_1}^{t_2}\int_{\Omega} \lambda\psi
    \,\ed t\, \ed x\,\ed y - \int_{t_1}^{t_2}\int_{\del \Omega} n_k j^k\,\ed t
    \,\ed S\,.
  \end{align}

  The boundary term in \eqref{eq:deviation_2} can be controlled by
  choosing appropriate boundary conditions. Inserting the definition
  of $j^k$ we find for the integrand
  \begin{align}\label{eq:deviation_3}
     n_k j^k \overset{\eqref{eq:momentum_flux_density}}
    {=}  \tderiv{\efv}^A\,n^k\del_k \efv_A \,.
  \end{align}
  In our numerical code we use homogeneous Neumann boundary conditions on all 
  the unknowns~$U_A$
  \begin{align*}
    n^k \del_k \efv_A = 0\,.
  \end{align*}
  With these boundary conditions the boundary term \eqref{eq:deviation_3} 
  vanishes and \eqref{eq:deviation_2} reduces to
  \begin{align*}
     \Delta E = \int_{t_1}^{t_2}\int_{\Omega} \lambda\psi\,\ed t\, \ed x\,\ed y
     \,.
  \end{align*}
  This is the change of the energy, which is due to the violation of
  the constraints.
  With other boundary conditions, for example Sommerfeld conditions, the 
  boundary term would remain and describe an energy flux through the boundary.
  \section{Numerical Setup}\label{sec:numerics}
  \subsection{Spatial Discretisation}
  In our investigations we use the Method of Lines approach for numerically 
  solving PDEs. This means here we discretized the spatial variables with 
  Finite Differences and evolve the resulting grid functions with either the 
  fourth order Runge-Kutta or the Rattle method.
  
  \begin{figure}[ht]
     \centering
     \includegraphics{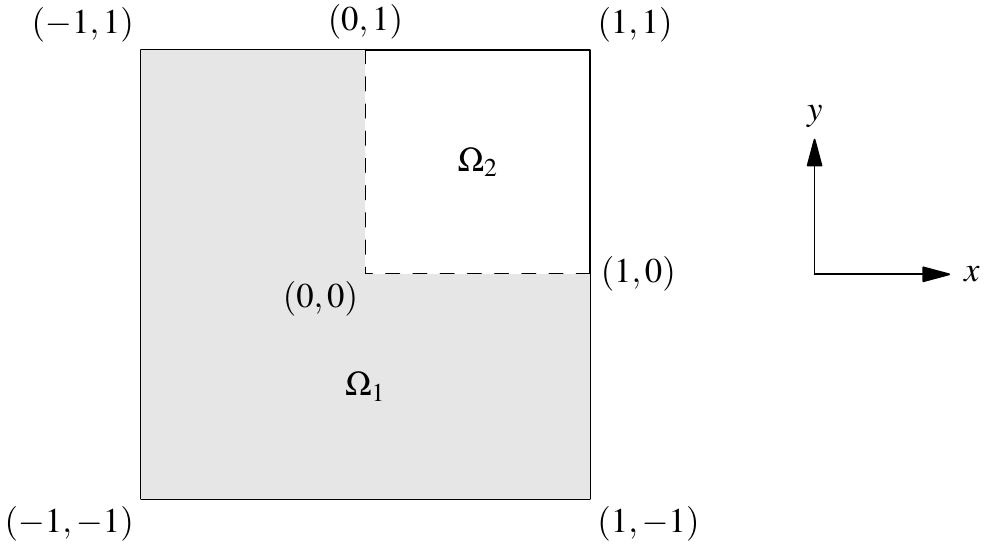}
     \caption{Domains of integration. Full domain $\Omega_1 = 
     [-1,1]\times[-1,1]$ and the reduced (quarter) domain $
     \Omega_2 = [0,1]\times[0,1]$.}
     \label{fig:0}
  \end{figure}
  
  In section \ref{sec:results_constr}, \ref{sec:results_origin} 
  and \ref{sec:results_energy} we are going to compare the results for 
  the time integration methods directly. There, we perform our simulations on 
  the  domain of integration $\Omega_1 = [-1,1]\times[-1,1]$ (see figure \ref
  {fig:0}) and discretize it via an equidistant grid.
  \begin{align*}
    x \longrightarrow x_j &= -1 + (j-1) h_x \qquad j = 1,...,N_x\\
    y \longrightarrow y_k &= -1 + (k-1) h_y \qquad k = 1,...,N_y\,.
  \end{align*}
  with the grid spacings $h_x = x_{j+1} - x_j$ and $h_y = y_{k+1} - y_k$. 
  All functions $f(x,y)$ will be evaluated on the grid nodes:
  \begin{align*}
    f(x,y) \longrightarrow f(x_j,y_k) = f_{j,k}\,.
  \end{align*}
  This leads to approximations of the derivatives by differences of the 
  function values at the grid points. For fourth order accuracy we write 
  the following standard five point formulas for the derivative with respect to 
  $x$ at a point $(j,k)$: 
  \begin{align*}
     \left.\xderiv{f}\right |_{(j,k)} &= \frac{f_{j-2,k} - 
     8f_{j-1,k} + 8f_{j+1,k} - f_{j+2,k}}{12 h_x},\\
     \left.\xxderiv{f}\right |_{(j,k)} &= \frac{-f_{j-2,k} 
     + 16f_{j-1,k} - 30f_{j,k} + 16f_{j+1,k} - f_{j+2,k}}{12 h_x^2}\,.
  \end{align*}
  Derivatives with respect to $y$ can be done equivalently. In all our 
  computations we use the same grid spacing: $h_x = h_y$ in $x$ and
  $y$ directions.
  
  For the simulations on the blow up dynamics in section \ref
  {sec:results_blowup}, we use the symmetry of the system to reduce the domain 
  of integration to a quarter of $\Omega_1$ only. This has the advantage, that 
  we effectively double the number of grid points with the same numerical 
  costs. The new grid covers therefore the domain $\Omega_2 = [0,1]\times[0,1]$ 
  (see also figure \ref{fig:0}).
  \subsection{Boundary Conditions}
  As mentioned above, we use homogeneous Neumann conditions
  for the numerical calculations where we compare the Runge-
  Kutta and the Rattle method (this is important for the above described energy 
  correction). We implement this boundary condition
  by imposing symmetries for the grid functions across the grid 
  boundary. This gives us the possibility to determine the grid functions on 
  points, which are beyond the boundaries (ghost points) which are
  needed for the evaluation of the finite difference operators. On the
  left boundary ($j = 1$) we get
  \begin{align*}
    f_{0,k} = f_{2,k} \quad \text{and} \quad f_{-1,k} = f_{3,k}
  \end{align*}
  and on the right boundary ($j = N_x$)
  \begin{align*}
    f_{N_x + 1,k} = f_{N_x - 1,k} \quad \text{and} \quad
    f_{N_x + 2,k} = f_{N_x - 2,k}
  \end{align*}
  and similarly we set the boundary conditions in the $y$ direction.
  
  The symmetries of the functions $u$ and $v$ make it necessary 
  to impose different boundary conditions for the computations on the reduced 
  domain $\Omega_2$. For the function $u$, we need along the boundary $x=0$ and 
  for the function $v$ along $y=0$ Dirichlet boundary conditions and not 
  Neumann ones. On those boundaries, we have $u(0,y) = 0$ respectively $v(x,0) 
  = 0$. On the remaining three boundaries we can keep the Neumann conditions. 
  To compute the ghost points for the Dirichlet boundary conditions we use the 
  following relations on the left boundary
  \begin{align*}
    f_{0,k} = -f_{2,k} \quad \text{and} \quad f_{-1,k} = -f_{3,k}
  \end{align*}
  The boundaries where we have to change the boundary conditions are for both 
  functions left boundaries, therefore, we not need the equivalent relations on 
  the right boundaries.
  \subsection{Initial data}
  The initial data we use are a polynomial bump with zeros at $r_1$ 
  and $r_2$:
  \begin{equation}\label{eq:ID_poly_ring}
  \begin{split}
    \vt_{0}(r) &=
    \begin{cases}
      A\,\left( 4\frac{(r - r_1)(r_2-r)}{(r_2 - r_1)^2} \right)^n
      &\text{for}\,\,r\in[r_1,r_2]\\
      0 &\text{otherwise.}
    \end{cases} 
  \end{split}
  \end{equation}
  The constant $A$ represents the amplitude of $\vt_{0}(r)$. By using the 
  relations \eqref{eq:trafo_intr_extr_u}, \eqref{eq:trafo_intr_extr_v} and 
  \eqref{eq:trafo_intr_extr_w} we get the initial data for the extrinsic 
  formulation.Throughout this article we use in all computations the parameters 
  $r_1 = 0.5$, $r_2 = 1$ and $n = 4$. The initial data \eqref{eq:ID_poly_ring} 
  describe a ring shaped bump in the $xy$-plane for the function
  $w(t,x,y)$ centred around the origin.
  
  In order that the Cauchy problem is well posed we also need initial data for 
  the velocities. We choose 
  \begin{align*}
    \tderiv{u}(0,x,y) &= \rderiv{u} = \cos \vt_0(r) \,\vt_0'(r) \,\frac{x}{r}\\
    \tderiv{v}(0,x,y) &= \rderiv{v} = \cos \vt_0(r) \,\vt_0'(r) \,\frac{y}{r}\\
    \dot{w}(0,x,y) &= \rderiv{w} = -\sin \vt_0(r) \,\vt_0'(r)\,.\phantom{\frac
    {x}{r}}
  \end{align*}

  With this choice of initial data for the velocities, the above described ring 
  shrinks towards the origin. This means that the energy concentrates around 
  the origin. After the function $w(t,x,y)$ has reached a minimum close to the 
  origin the ring starts expanding again. If the energy concentration
  at the origin is 
  large enough, one expects a singularity formation like it is assumed in the 
  three conjectures \ref{conj:Bizon_1}, \ref{conj:Bizon_2}, \ref
  {conj:Bizon_3} above.
  
  Rapha\"el and Rodnianski presented in 
  \cite{RaphaelRodnianski} a set of initial data for analytical, but also 
  numerical investigations. However we have chosen the initial data \eqref
  {eq:ID_poly_ring} to compare our \textit{numerical} results with the ones in 
  \cite{Bizon2001}.
  
%
  \section{Results}\label{sec:results}
  The main aim in this article is to compare the evolution of a constrained 
  system by using an integration method for a free evolution and a method, 
  which explicitly takes the constraint equations into account. As our model 
  we choose the above described 2+1 dimensional Wave Map equations in the 
  extrinsic formulation, as an example for a wave-like PDE system, which is 
  constrained by additional (algebraic) conditions.
  
  In the case of the free evolution we use the equations \eqref
  {eq:extr_eom_u}, \eqref{eq:extr_eom_v}, \eqref{eq:extr_eom_w} and replace the 
  Lagrangian multiplier $\lambda$ with the expression \eqref{eq:lag_mult_uvw}. 
  For this calculation we use the standard fourth order Runge-Kutta 
  method (RK4). These results will be compared with the ones gained from the 
  time evolution taking the constraint equation into account. Here we are going 
  to use equations \eqref{eq:extr_eom_u}, \eqref{eq:extr_eom_v}, \eqref
  {eq:extr_eom_w}, \eqref{eq:extr_constr_uvw} and \eqref
  {eq:extr_velo_constr_uvw}. The time evolution will be done with the Rattle 
  method (RTL) described in section \ref{sec:Rattle}. For the Rattle method we 
  choose for the maximally allowed constraint violation the value   
  $10^{-12}$. This means that for all times $t$: $\|\phi\|_{\text{max}}(t) \leq 
  10^{-12}$ as well as $\|\psi\|_{\text{max}}(t) \leq 10^{-12}$, where $\|\cdot
  \|_{\text{max}}$ is the maximum norm. If this accuracy cannot be reached 
  within 100 iterations the program aborts. For the computations presented here 
  this was never the case. All following computations are done in the time 
  interval $t\in[0,1.6]$ and with a Courant-Friedrichs-Lewy factor $\Delta t/h 
  = 0.2$. In the simulations, where we compare the two integration methods, we 
  have chosen the values for the amplitude $A$ in a way that none of the two 
  integration methods breaks down.
  \subsection{Constraint functions}\label{sec:results_constr}
  One of the main points in this article is the preservation of the constraint 
  \eqref{eq:extr_constr_uvw} and its derivative \eqref
  {eq:extr_velo_constr_uvw}. In figure \ref{fig:1} we compare the constraint 
  violation for the two different integration methods. We see clearly that the 
  Runge-Kutta method leads to a permanent increase of the constraint violation 
  during the time evolution. The Rattle method always keeps the constraint 
  violation below the previously set value $10^{-12}$.
  
  For the velocity constraint $\psi$ we can see in figure \ref{fig:2} 
  qualitatively the same results as for the constraint $\phi$. But it is worth 
  mentioning that the Rattle methods preserves the velocity constraint even a 
  few orders better than requested.
  
  The accelerated increase of the constraint violation for the Runge-Kutta 
  method at $t \approx 0.8$ occurs at the moment when the wave packet is 
  reflected at the origin. During this reflection process the spatial 
  derivatives near the origin are very high. This also leads to a strong 
  deviation of the function from its theoretically fixed value at the origin.
  
  \begin{figure}[ht]
     \centering
     \includegraphics{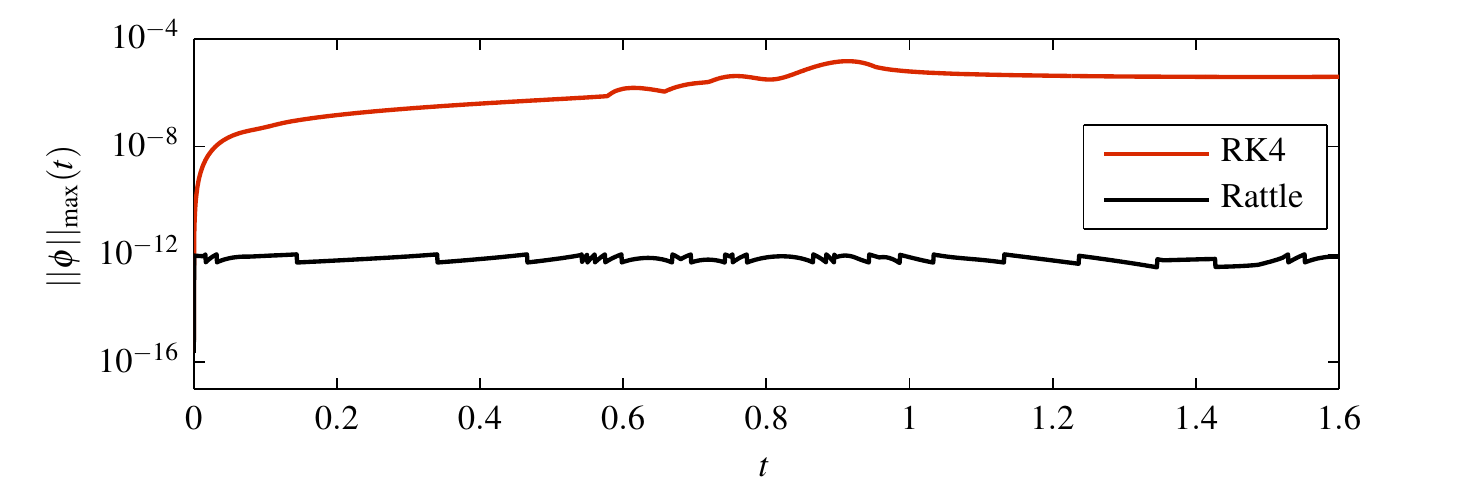}
     \caption{Maximum norm $||\phi||_{\text{max}}(t)$ for the constraint $\phi$ 
     during the time evolution for $A = 0.4$.}\label{fig:1}
  \end{figure}
  
  \begin{figure}[ht]
     \centering
     \includegraphics{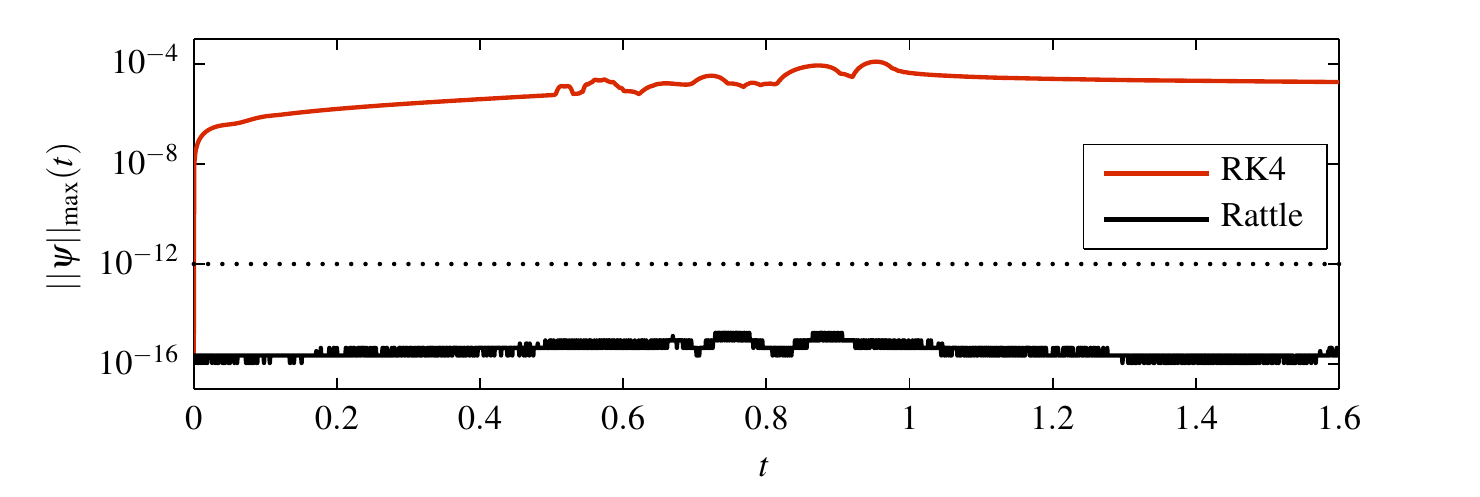}
     \caption{Maximum norm $||\psi||_{\text{max}}(t)$ for the velocity 
     constraint $\psi$ during the time evolution for $A = 0.4$.
     The dotted line is the preset value for the maximum of 
     the constraint violation.}
     \label{fig:2}
  \end{figure}
  
  We mention here, that the convergence for both methods is of 
  the expected order. If we rescale the results shown in figure \ref{fig:1} and 
  \ref{fig:2} for different numerical resolutions with the expected, fourth 
  order, convergence 
  rate, the curves for the Runge-Kutta method cannot visually be 
  distinguished. To check the convergence of the Rattle method, we cannot use 
  the constraint preservation, because in this method we predefine the maximum 
  of the constraint violation.
  However, by checking the convergence rate with the function $w$
  with the function $w$ we also obtain the expected second order convergence.
  \subsection{Behaviour at the Origin}\label{sec:results_origin}
  At the origin $r=0$, respectively $(x=0,y=0)$ the solution $\vt(t,r)$ of the 
  equivariant Wave Map equation \eqref{eq:equiv_wm} has the noteworthy feature 
  that it is constant during the whole time evolution. One can 
  show that
  from the equations that the solution evolving from our class of
  initial data necessarily satisfies for all $t$
  \begin{align*}
    \vt(t,0) = 0 \qquad \text{and respectively} \qquad w(t,0,0) = 1\,.
  \end{align*}
  This condition is not enforced in our code.
  In figure \ref{fig:3} we compare how the two integration 
  methods preserve this property. We can see that the Rattle method always 
  keeps the deviation $|w(t,0,0) - 1|$ under the predefined maximum of $10^
  {-12}$ for the constraint violation. As expected from the previous results, 
  the Runge-Kutta method cannot preserve the property, because it lets
  the solution drift away from the constraint manifold. In the 
  literature, this 
  drift is a well known phenomenon for constrained ODEs (see \cite
  {LeimkuhlerReich}). Numerical solutions of constrained systems, in the 
  formulation where the Lagrangian multipliers are eliminated, show this 
  behaviour in general.
    
  For larger values  of $A$ the Runge-Kutta method breaks down at the origin. 
  This breakdown depends on the spatial resolution and is therefore a numerical 
  phenomenon and not an indication for the collapse of the solution. The 
  solution gained with the Rattle method shows a different behaviour. For large 
  values of $A$ the solution flips from $w(t,0,0) = 1$ to $w(t,0,0) = -1$. This
  indicates that it becomes increasingly difficult for the iterative
  algorithm of the constraint equation to find a solution and
  ultimately the iteration seems to approach the other static solution,
  the stereographic projection from the north pole, for which $w(t,0,0)
  = -1$. We take this behaviour as an indication that the solution is
  in the blow-up regime. For the critical value of the amplitude, we
  take the value where the behaviour at the origin changes and find $A^{*} 
  \approx 0.81871173$.

  \begin{figure}[ht]
     \centering
     \includegraphics{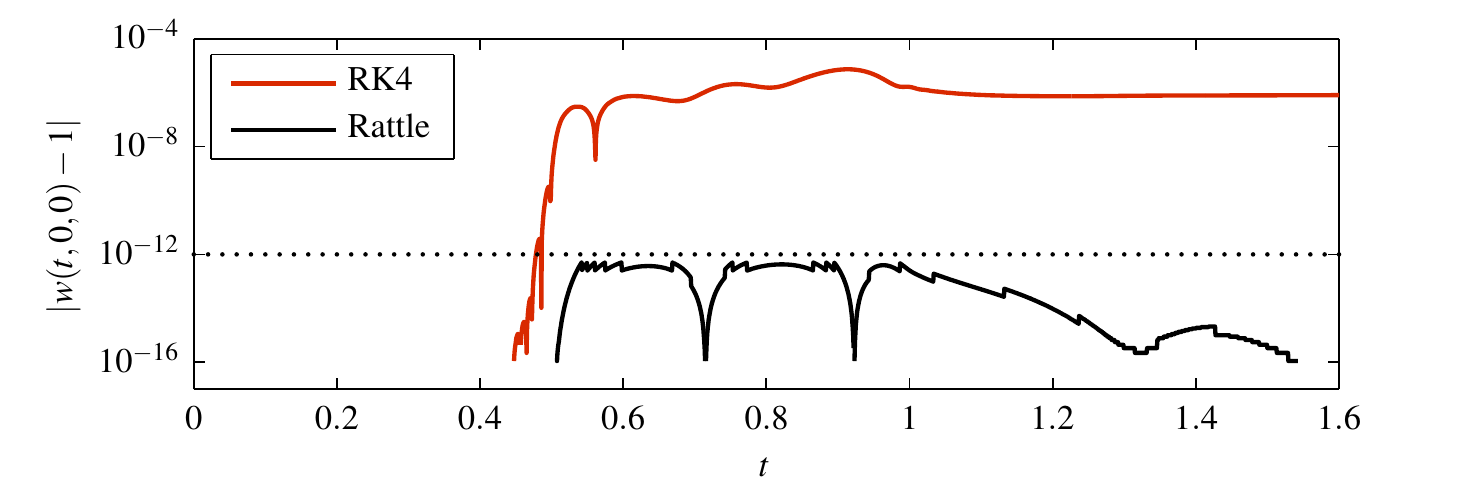}
     \caption{Displacement $|w(t,0,0) - 1|$ of the solution at the origin for 
     $A = 0.4$ during the time evolution. The dotted line is 
     the preset value for the maximum of the constraint 
     violation.}\label{fig:3}
  \end{figure}
  
%
  \subsection{Energy Conservation}\label{sec:results_energy}
  An often used measure for the quality of numerical solutions of ODEs and PDEs 
  is the energy conservation during the time evolution. Of course, this makes 
  sense for non-dissipative systems only. We compare the energy 
  conservation for the evolution computed with the Runge-Kutta and the ones 
  obtained with the Rattle method. We can see in figure \ref{fig:4} that the 
  average energy conservation with the Rattle method is better than with the 
  Runge-Kutta method.

  In section \ref{sec:flux} we presented the analytical formula for an energy 
  loss, which is caused by the violation of the constraint
  equations. Our numerical investigations confirm the existence of this energy 
  loss. We can use this term to correct the total energy $E_{\text{corr}} = E - 
  \Delta E$. From figure \ref{fig:4} we see that the corrected energy is better 
  preserved than without correction, and also in average better than with the 
  Rattle method. Only during the reflection at the origin the Rattle method 
  results in an better energy conservation.
  
  The energy correction computed for the Rattle method is so small compared to 
  the relative energy, that it can be neglected.
  
  \begin{figure}[ht]
     \centering
     \includegraphics{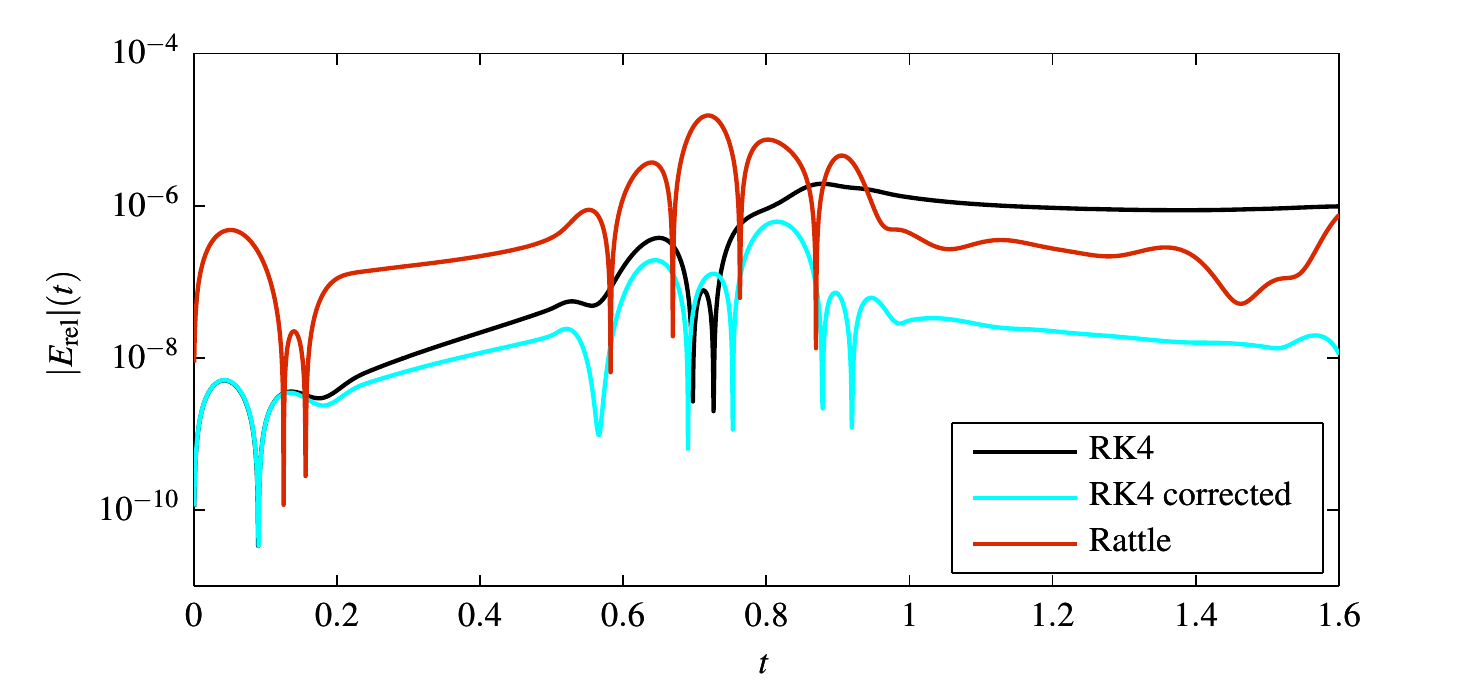}
     \caption{Relative energy $|E_{\text{rel}}|(t) = \left| 1 - E(t)/E
     (0)\right|$ computed with the Runge-Kutta (corrected and uncorrected) 
     and the Rattle methods for $A = 0.4$ and $N = 641$.}\label{fig:4}
  \end{figure}
  \subsection{Scaling Function}\label{sec:results_blowup}
  The conjectures by Bizo\'n et.al. refer to the scaling function $s
  (t)$, which we determined in \eqref{eq:scaling_func}. This function
  plays a central role in the study of the singularity formation of
  Wave Maps \cites{OvchinnikovSigal,RaphaelRodnianski}. The importance
  comes from the fact, that with the help of $s(t)$ the dynamical
  solution can be rescaled in a way that it approximates the static
  solution during the singularity formation, around the point where
  the singularity will appear. This means the scaling function is
  directly involved in the blow-up. The fact that the scaling 
  function plays an crucial role in the singularity formation was shown by 
  Struwe in \cite{Struwe2003_2}. The scaling function is an direct indicator 
  for the shrinking of the static solution, which drives the singularity 
  formation.

  In this part we determine the scaling function for different values
  of the criticality parameter $A$ in its dependence on time. All the
  calculations in this subsection have been performed with the Rattle
  method because of its superior behaviour in the critical regime.
  As mentioned before the simulations for this part were computed on the 
  reduced domain of integration $\Omega_2$.
  
  As we can see in figure \ref{fig:ScalFunc}, increasing the
  amplitude $A$ towards the critical value $A^{*}$, the time interval
  in which the scaling function is approximately constant,
  increases. This means, that during this period the solution remains
  almost constant, hovering in a quasi-static state.
  But for higher resolutions 
  this states last shorter. Therefore we are very confident the method we 
  presented is able to deal with such critical situations, but the need for 
  much higher resolutions is evident. We interpret this quasi-static state as 
  the fact that the numerical scheme cannot follow the shrinking of the 
  harmonic map anymore. Therefore for this numerical resolution the critical 
  situation occurred. When the scaling function increases again, the resolution 
  is again sufficient to follow this process.
  \begin{figure}[ht]
     \centering
     \includegraphics{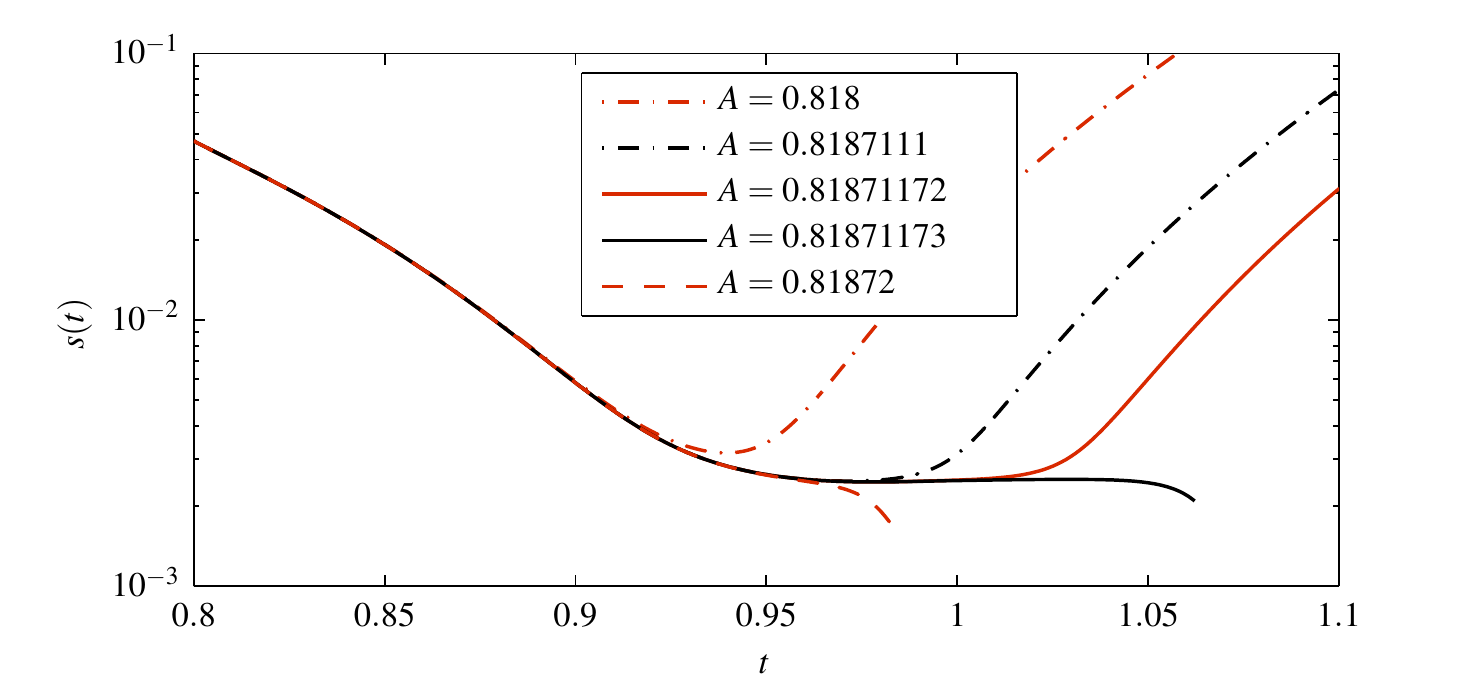}
     \caption{The scaling function $s(t)$ for different parameter values $A$. 
     The value $A^* = 0.81871173$ is the critical amplitude.}\label
     {fig:ScalFunc}
  \end{figure}
  
  In figure \ref{fig:5} we can see the scaling function $s(t)$ for
  different values of the amplitude parameter $A$. The black 
  curve is the fit to the curve for the amplitude $A = 0.81871172$, which is 
  the one closest to the critical value $A^*$. The interval $t\in[0.836,0.85]$ 
  between the dotted lines is the domain used for the fitting procedure. 
  The fit was done with respect to the analytic expression
  \begin{align*}
    s(t) = \frac{1.04}{\expo}\,(T-t)\,\exp\left(-\sqrt{-\ln\left(T - t\right) + 
    b}
    \right)
  \end{align*}
  which is taken from \cite{OvchinnikovSigal}. The parameter $b$ depends on the 
  choice of the initial data. Our choice for the interval for the fit is a 
  compromise between being close enough to the blow up time and the domain 
  where we can be sure, that the scaling function numerically converges. The 
  results presented here were computed with $N = 641$ grid points in $x$ and $y
  $ direction. For higher resolutions the amplitude $A^* = 0.81871173$ is not 
  critical at all.
  The results of the fit were $T = 0.94151363$ and $b = -2.02978334$ with a 
  residual error of $1.31486579\cdot10^{-8}$. The fit parameter were computed 
  with the \textsc{Matlab} function \texttt{lsqcurvefit} from the Optimization 
  Toolbox.
   
  \begin{figure}[ht]
     \centering
     \includegraphics{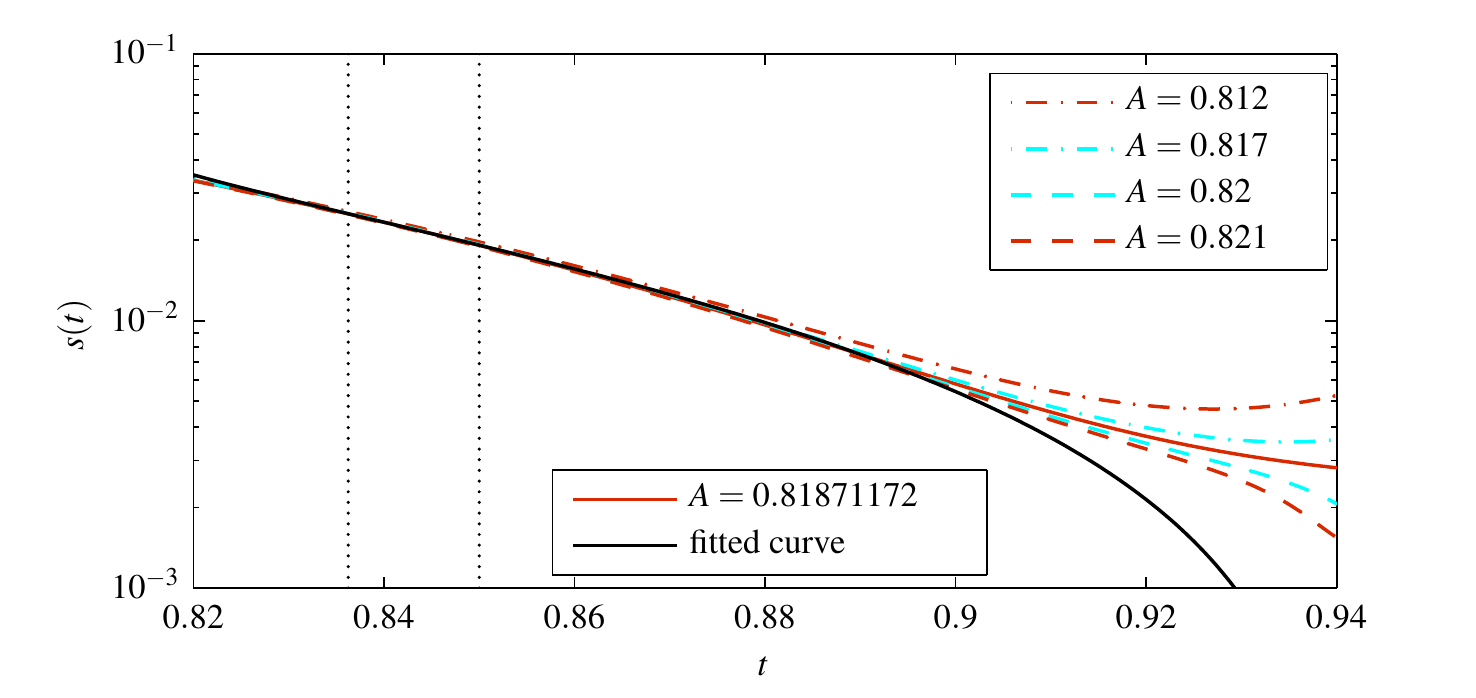}
     \caption{Fit of the scaling function $s(t)$ to the analytical expression, 
       given in \cite{OvchinnikovSigal}. The computed blow up is $T = 
       0.94151363$. The value $A = 0.81871172$ is the last
       under-critical value shown, i.e., for which the solution does
       not change its behaviour at the origin.}\label{fig:5}
  \end{figure}
  
  Figure \ref{fig:6} shows the rescaled function $w(t,r/s(t))$ at
  different times and the static solution $w_{\text{S}}(r)$. The
  amplitude is $A = 0.81871172$, very close to the critical value
  $A^{*} = 0.81871173$. At $t = 0.919375$ the solution reaches its
  overall minimum value of $w = -0.91372693$. Before this time
  the wave packet moves towards the origin, where it is reflected and
  subsequently moves away from the origin. We can see that the
  rescaled dynamic solution is approximated very well by the static
  solution near the origin. As mentioned before for larger amplitudes
  the solutions show a flip of the solution at the origin. In the
  caption of figure 5 in \cite {Bizon2001}, Bizo\'n et. al. note the
  fact that the solution overshoots the value $-\pi$, which they
  consider as a necessary and sufficient condition for the blow up. We believe
  that in our case it is the observed flip at the origin, which
  indicates the same phenomenon.
  
  \begin{figure}[ht]
     \centering
     \includegraphics{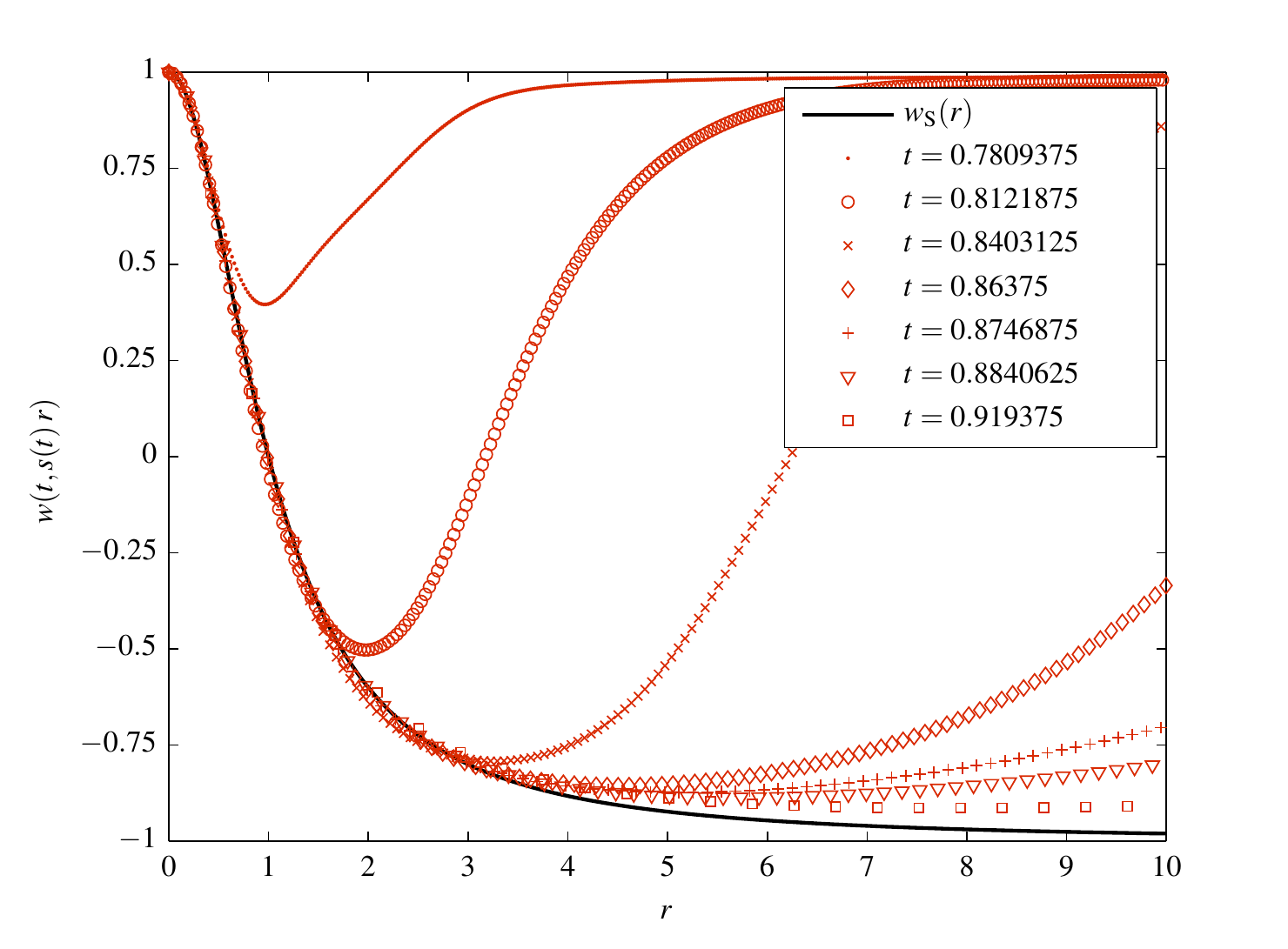}
     \caption{Ingoing wave packet for $A = 0.81871172$ rescaled with
       the scaling function $s(t)$ for various values of $t$. The
       rescaled functions $w(t,s(t) r)$ approximate the static
       solution $w_{\text{S}}(r)$. The function $w(t,r)$ reaches its
       minimum at $t = 0.919375$. The computation was
       done with $N = 641$ grid points on the quarter grid.}
     \label{fig:6}
  \end{figure}
  %
  %
%
  \section{Conclusion}
  In this article we demonstrated that the Rattle method can be very useful for 
  the numerical time evolution of relativistic field equations. In terms of 
  constraint conservation it is, by construction, superior compared to a free 
  evolution scheme like the standard fourth order Runge-Kutta method. The usage 
  of the constraint equations also leads to a more accurate behaviour in 
  extreme situations, like we showed it for the behaviour of the Wave Map 
  solution at the origin. At this point the solution should always take the 
  same value. The Rattle method preserves this symmetry in the accuracy 
  previously set as the limit of the constraint violation. The Runge-Kutta 
  method leads to a drift away from this state.
  
  We presented an energy correction term, which is a direct consequence of the 
  constraint violation. The explicit computation of this deviation was used to 
  correct the total energy. This energy deviation depends on the magnitude of 
  the constraint violation, which is the reason why it can be ignored with the 
  Rattle method by choosing the allowed constraint violation small enough. For 
  the Runge-Kutta method it was possible to correct the energy in a way that 
  the energy conservation during the whole time evolution is better than for 
  the Rattle method. But this better energy conservation does not cure the 
  other shortcomings namely the increasing constraint violation and violation 
  of the symmetry at the origin.
  
  Finally we presented some results obtained in the equivariant case
  using the Rattle method. We computed the scaling function $s(t)$ for
  different values for the amplitude and confirmed the expected
  behaviour that the rescaled dynamic solution approximates the static
  solution as the parameter $A$ approaches a critical value. Since we
  obtain this behaviour without imposing the equivariance explicitly,
  this shows that the critical behaviour observed in the 1+1 codes is
  in fact stable under small non-spherical perturbations, which are
  inadvertently introduced due to the numerical errors. To our
  knowledge it has not been shown numerically that critical behaviour
  will also appear in the non-equivariant case. In a future 
  article, we are going to show that under explicit violation of the 
  equivariance, we still observe the critical behaviour.
  
  
  It is obvious from our results, that for an insight into the 
  delicate 
  details of the blow up dynamics much higher numerical resolutions are 
  required. These could be achieved by using grid refinement techniques for 
  example. However, the Rattle method applied to the extrinsic formulation of 
  the 2+1 Wave Map seems to be a promising way to study this system without 
  running into problems with coordinate singularities.

%
%

%
%
\end{document}